\documentclass[prc,reprint,superscriptaddress,nofootinbib]{revtex4-1}

\usepackage{graphicx}
\usepackage{color}
\usepackage{units}
\usepackage{xcolor,colortbl}
\definecolor{prediction}{rgb}{0.9,0.9,0.9}
\usepackage{amsmath,amssymb,bm}
\usepackage{amsfonts}
\usepackage[latin1]{inputenc}
\usepackage[caption=false]{subfig}
\usepackage[colorlinks,urlcolor={blue}]{hyperref}

\usepackage{todonotes}

\newcommand{\co}{}

  \def\nuc#1#2{\relax\ifmmode{}^{#1}{\protect\text{#2}}\else${}^{#1}$#2\fi}
  \def\itnuc#1#2{\setbox\@tempboxa=\hbox{\scriptsize\it #1}
    \def\@tempa{{}^{\box\@tempboxa}\!\protect\text{\it #2}}\relax
    \ifmmode \@tempa \else $\@tempa$\fi}

\newcommand{\xEFT}{\ensuremath{\chi}EFT}

\newcommand{\OEd}{\ensuremath{E({}^2\text{H})}}

\newcommand{\OEa}{\ensuremath{E({}^4\text{He})}}

\newcommand{\Orpd}[1][{}]{\ensuremath{r_{\text{pt-p}}^{#1}({}^2\text{H})}}

\newcommand{\Orpa}[1][{}]{\ensuremath{r_{\text{pt-p}}^{#1}({}^4\text{He})}}

\newcommand{\LECvec}{\ensuremath{\bm{\alpha}}}

\newcommand{\LEC}{\ensuremath{\alpha}}

\DeclareMathOperator{\tr}{tr}

\graphicspath{{fig/}}

\begin{document}

\title{Quantifying statistical uncertainties in ab initio nuclear physics using Lagrange multipliers} 

\author{B.~D.~Carlsson} \email{borisc@chalmers.se}
\affiliation{Department of Physics,
  Chalmers University of Technology, SE-412 96 G\"oteborg, Sweden}

\date{\today}

\begin{abstract}
Theoretical predictions need quantified uncertainties
for a meaningful comparison to experimental results.
This is an idea which presently permeates the field of theoretical nuclear physics.
In light of the recent progress in estimating theoretical uncertainties in
ab initio nuclear physics, we here present and compare methods for evaluating
the statistical part of the uncertainties.
A special focus is put on the (for the field) novel method of Lagrange multipliers (LM).
Uncertainties from the fit of the nuclear interaction to experimental data are propagated
to a few observables in light-mass nuclei to highlight any differences between the presented methods.
The main conclusion is that the LM method is more robust, while covariance based methods are less
demanding in their evaluation.
\end{abstract}

\maketitle

\section{Introduction\label{sec:introduction}} 

Although it is crucial for experimental results to have estimated uncertainties,
the same has not always been acknowledged for theoretical calculations and predictions.
This is, however, beginning to change. A recent editorial for the journal Physical Review A
highlights the importance of estimated uncertainties for theoretical calculations~\cite{editors2011}
and urges authors to include such estimates.
Practitioners are starting to pay more attention to this important aspect~\cite{dudek2013,chen2014,lu2016}.
It is being acknowledged that it is important to understand all sources of uncertainties in the theory, model
and numerical calculations. Not only is this important for a meaningful comparison to experimental data,
it also has the potential benefit of increasing the understanding and awareness of missing physics in the model.
Therefore, the quantification of theoretical uncertainties in low-energy nuclear physics,
and chiral effective-field theory (\xEFT) in particular,
has recently received much attention.
Various methods and strategies are being introduced to the field,
such as statistical sensitivity analyses~\cite{dobaczewski2014,carlsson2016},
strategies for estimating model uncertainties~\cite{epelbaum2015,carlsson2016},
including Bayesian methods~\cite{furnstahl2015,furnstahl2015b,wesolowski2016},
and advanced statistical tools~\cite{navarro2014}.
Statistical error propagation has been performed using
various methods~\cite{ekstroem2015,navarro2015b,carlsson2016} and is
being expanded to more and more nuclear observables~\cite{navarro2015d,carlsson2016,acharya2016}.

Due to the increasing demand for nuclear interaction models with quantified uncertainties,
it is of interest to compare different methods for extracting uncertainties.
Such a comparison serves several purposes:
\begin{itemize}
	\item Investigate if, and if so why, different methods produce different results.
	\item Justify or reject various approximations involved in these methods.
	\item To serve as a reference and guide for future works intending to use these methods.
\end{itemize}

The process of going from a \xEFT{} interaction to a predicted value for an observable
involves many sources of uncertainties.
There is an inherent \emph{model error} in \xEFT{} due to the exclusion of higher-order terms in the
interaction. In the solution of the many-nucleon Schr\"{o}dinger equation,
there can be a sizable \emph{method error} from e.g.\ truncation of the number of particle-hole excitations
and limited model spaces,
and sometimes also a \emph{numerical error} due to round-off errors.
Finally, there is a \emph{statistical uncertainty} from e.g.\ the fitting of the low-energy constants (LECs) to
experimental data. The LECs determine the strength of contact interactions in the chiral Lagrangian,
which are not fixed by chiral symmetry.
The numerical values of the LECs are determined by choosing the LECs that best describe
a set of experimentally measured observables.
Since experimental data comes with uncertainties,
this fitting results in statistical uncertainties in the LECs,
or rather a multi-variate probability distribution for the values of the LECs.
This probability density is then propagated to observables to yield a statistical uncertainty.
To compensate for a lack of data and avoid over-fitting, \emph{priors}
for the LECs can be applied~\cite{furnstahl2015,furnstahl2015b,wesolowski2016}.
The priors incorporate \emph{a priori} knowledge of the model from e.g.\ the underlying theory
to constrain the LECs to feasible values.
However, such an approach will not be investigated here.

This article will focus on the statistical uncertainties and how they are quantified.
It has been shown that these uncertainties are generally small, compared
to the model uncertainties of \xEFT~\cite{epelbaum2015,carlsson2016}.
Nevertheless, statistical covariances carry useful information; they can be used
to study correlations between observables and for doing sensitivity analyses to determine what
experimental data could be used to
constrain other observables further~\cite{reinhard2010,kortelainen2010,piekarewicz2015}.

The purpose of this article is to see how some common methods used to propagate statistical uncertainties
relate and how they compare, both with regard to actual values for the statistical uncertainties
but also in their ease of application and computational requirements.
Special attention will be given to the method of
Lagrange Multipliers~\cite{stump2001} (LM), as this method has, to the author's knowledge,
not been used in \xEFT{} studies before.

In Sec.~\ref{sec:method} the different methods for extracting statistical uncertainties are presented.
In Sec.~\ref{sec:results} the obtained uncertainties are compared for some observables.
Finally, in Sec.~\ref{sec:discussion}, there are some concluding remarks about the methods.

\section{Method}\label{sec:method}

As mentioned in Sec.~\ref{sec:introduction}, the statistical uncertainties originate from
the experimental uncertainties through a fit of the LECs to data.
The standard method to find the \emph{optimal} LECs $\LECvec_0$ in \xEFT{} is to
perform a non-linear least-squares minimization~\cite{dobaczewski2014}, of the general form
\begin{align}\label{eq:chi2}
	\chi^2(\LECvec) = \sum_{n=1}^N \left(
		\frac{O_n^{\rm (exp)} - O_n^{\rm (theo)}(\LECvec)}{\sigma_n^{\rm (tot)}}\right)^2
	\equiv \sum_{n=1}^N r_n^2(\LECvec).
\end{align}
Here, $O_n^{\rm (exp)}$ is the experimental value for observable $n$,
$O_n^{\rm (theo)}$ is the corresponding theoretical prediction, $\sigma_n^{\rm (tot)}$ is the combined
uncertainty of the experimental and theoretical value and
$r_n$ are known as \emph{residuals}.
This minimization yields the value $\chi_0^2 \equiv \chi^2(\LECvec_0)$.
One benefit of using Eq.~\eqref{eq:chi2} is that it has well-known statistical properties,
under certain conditions, allowing for
the propagation of the experimental uncertainties to the LECs~\cite{dobaczewski2014}.

The basis for the statistical analysis is that $\chi^2(\LECvec)$ follows a chi-squared distribution with
$N_{\rm dof} = N - N_{\LECvec}$ degrees of freedom, where $N_{\LECvec}$ is the number of LECs.
This is the case if and only if all residuals $r_n$ are independent and normally distributed
with mean $0$ and variance $1$.
The range of variation allowed for the LECs, within one standard deviation,
is then given by all LECs $\LECvec$ that satisfy
$\Delta\chi^2(\LECvec) \equiv \chi^2(\LECvec) - \chi_0^2 < L = 1$~\cite{dobaczewski2014}.

In practice, the above conditions on the residuals are rarely completely fulfilled.
In particular, the presence of non-negligible systematic uncertainties can cause the residuals to
deviate from the normal distribution.
Underestimated or omitted systematic uncertainties can result in $\chi_0^2 > N_{\rm dof}$.
In this case, a global rescaling of the uncertainties with a so called Birge factor~\cite{birge1932}
can be applied, resulting in $L = \chi_0^2/N_{\rm dof}$.
This will make the variance of the residuals equal to unity.

Although these deviations in the distribution of the residuals
could compromise the statistical uncertainties~\cite{navarro2014},
it has been shown that they are stable despite
small deviations from normality~\cite{carlsson2016}.
This indicates that obtained uncertainties, correlations and sensitivity analyses still yield useful information.
However, further checks on the correctness of obtained statistical uncertainties are motivated
to make sure this is the case.

We will here present six methods, using different approximations and compare the results.
These methods can, in essence, be separated into two different strategies,
(i) Using the covariance matrix of the LECs to propagate uncertainties and
(ii) Use LM to obtain propagated uncertainties.

\subsection{Covariance matrix methods}

At the minimum defined by the LECs $\LECvec_0$, the Taylor expansion of $\chi^2(\LECvec)$ is given by
\begin{align}\label{eq:chi2_taylor}
    \chi^2(\LECvec_0 + \Delta\LECvec) \approx \chi_0^2 + \frac{1}{2}{(\Delta\LECvec)}^T H_0 \Delta\LECvec.
\end{align}
By construction, the Jacobian $J_0$ is zero in the minimum.
Furthermore, the elements of the Hessian matrix, $H_{0,ij}$, are given by
\begin{align}\label{eq:H_elem}
	H_{0,ij} = \sum_{n=1}^N \left(2\frac{\partial r_n}{\partial \LEC_i}\frac{\partial r_n}{\partial \LEC_j}
		+ 2r_n\frac{\partial^2 r_n}{\partial \LEC_i\partial \LEC_j}\middle)\right|_{\LECvec=\LECvec_0}.
\end{align}
In a computer implementation, the derivatives in \eqref{eq:H_elem} are typically obtained using either
finite differences or automatic differentiation. In the former case, to avoid the numerically difficult
task of computing second derivatives, an accurate approximation of $H_0$ is often used~\cite{press1992},
\begin{align}
	\tilde{H}_{0,ij} = \sum_{n=1}^N 2\left.\frac{\partial r_n}{\partial \LEC_i}\frac{\partial r_n}{\partial \LEC_j}\right|_{\LECvec=\LECvec_0}.
\end{align}
Since $\sum_{n=1}^N r_n \approx 0$, large cancellations can be expected to occur in
the omitted second-derivative term which justifies this approximation.

From the Hessian, or the curvature of the $\chi^2$ surface,
the covariance matrix for the LECs is given by
\begin{align}
	C &= 2LH_0^{-1}\\
	\tilde{C} &= 2L\tilde{H}_0^{-1},
\end{align}
with $L$ usually given by
\begin{align}\label{eq:L}
	L = \frac{\chi_0^2}{N_{\rm dof}}.
\end{align}
The probability distribution for the LECs are then given by
the multivariate normal distribution with central value $\LECvec_0$ and covariances $C$ or $\tilde{C}$.

There are various methods to propagate the statistical uncertainties to a general observable
$O(\LECvec)$. The most exact of the methods presented here,
is to perform a Monte Carlo sampling using $M$ samples,
resulting in the mean $\mu$ and variance $\sigma^2$ given by
\begin{align}
	\mu_{\rm sample} &= \frac{1}{M}\sum_{m=1}^M O(\LECvec_m)\\
	\sigma_{\rm sample}^2 &= \frac{1}{M-1}\sum_{m=1}^M \left(O(\LECvec_m) - \mu_{\rm sample}\right)^2,
\end{align}
where $\LECvec_m$ are sampled from the distribution of the LECs.
For an accurate result, a large number of samples is needed, often in the range $M \gtrsim 10^4 - 10^5$,
although much smaller sample sizes have been used also~\cite{navarro2015b,navarro2015d}.
In the results presented here, $M=10^5$ has been used.

Instead of performing a costly Monte Carlo sampling,
it is possible to use a Taylor expansion of the observable value around its central value,
\begin{align}\label{eq:O_approx}
	O(\LECvec_0 + \Delta\LECvec) \approx O_0 +
										 j_0^T\Delta\LECvec +
										 \frac{1}{2}(\Delta\LECvec)^Th_0\Delta\LECvec,
\end{align}
where $O_0$ is the value, $j_0$ is the gradient and $h_0$ the Hessian of $O$ at the point $\LECvec_0$.
From Eq.~\eqref{eq:O_approx}, a linear and a quadratic approximation to
the propagated statistical uncertainty are obtained,
\begin{align}
	\mu_{\rm linear} &= O_0\\
	\label{eq:sigma_linear}\sigma_{\rm linear}^2 &= j_0^T Cj_0\\
    \mu_{\rm quad.} &= \mu_{\rm linear} + \frac{1}{2}\tr\left(Ch_0\right)\\
    \sigma_{\rm quad.}^2 &= \sigma_{\rm linear}^2 + \frac{1}{2}\tr\left({(Ch_0)}^2\right).
\end{align}
The main benefit of the linear approximation is that no second derivatives are needed.
However, the calculation of the covariance matrix $C$ still involves second derivatives.
Therefore, I will also compare the results obtained using $\tilde{C}$,
\begin{align}
	\mu_{\rm approx.} &= \mu_{\rm linear}\\
	\sigma_{\rm approx.}^2 &= j_0^T\tilde{C}j_0.
\end{align}

The four methods based on the covariance matrix strategy described here
-- denoted \emph{sample}, \emph{quadratic}, \emph{linear} and \emph{approx} --
gradually employ more and more approximations, making it possible to track the source of possible deviations.

\subsection{Lagrange Multiplier methods}

\begin{figure}
    \centering
    \includegraphics[width=\columnwidth]{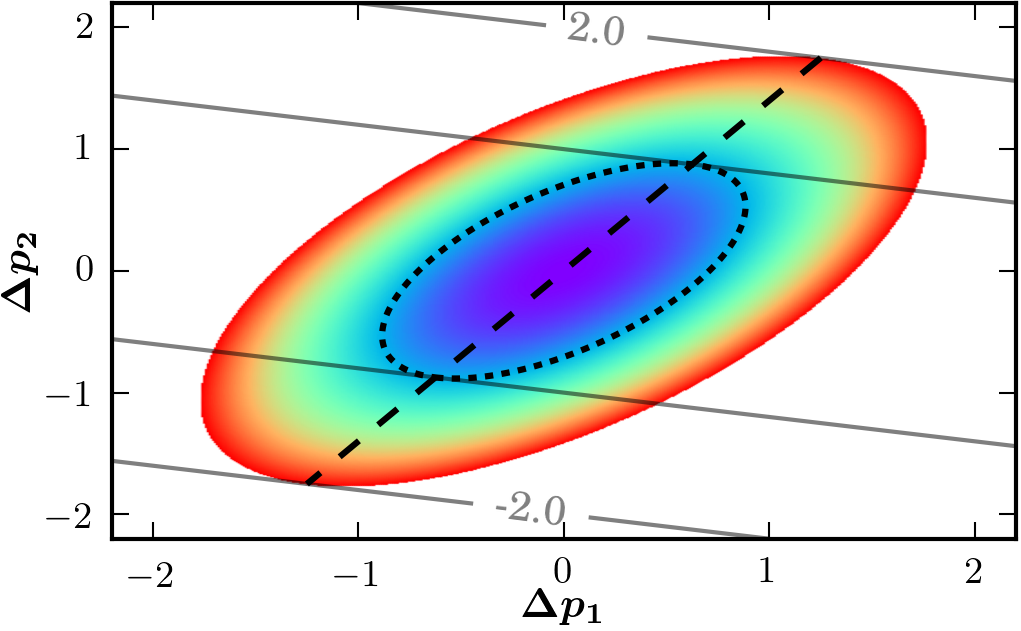}
    \caption{\label{fig:LM_example_linear}\co Illustration of the LM method.
    	A two-parameter chi-squared surface is shown as a filled surface, with the edge corresponding
    	to $\Delta\chi^2 = L$.
    	The solid lines are contour levels of an observable $O$ with central value $0$.
    	The statistical uncertainty of the LM method is given by all contour lines that crosses the
    	filled surface.
    	In this case this results in an uncertainty of $\pm 2$.
    	The dashed line represents the parameter values obtained for various fixed values of the
    	observable $O$.
    	The dotted line is a contour line for the chi-squared surface.
    	As expected, the dashed line crosses the contour lines of $O$ at the point where
    	the chi-squared value is lowest.}
\end{figure}

An alternative to using a covariance matrix is the LM method~\cite{brock2000,pumplin2001,stump2001}
where the statistical uncertainty of an observable $O$ is found through the use of
constrained minimizations.
The main advantages of the LM method is that it makes no assumption on the form of the chi-squared surface
around the minimum nor on the functional dependence of the observable on the LECs.
In the case of a quadratic chi-squared surface and a linear dependence of the observable $O$ on the LECs,
the LM method is equivalent to the linear method defined in the previous sections.
The statistical variability of $O$ is given by all values $O$ that
are attainable under the constraint that $\chi^2 \leq \chi_0^2 + L$,
with $L$ given in Eq.~\eqref{eq:L}.
The method is illustrated in Fig.~\ref{fig:LM_example_linear}.

To find the statistical uncertainty, minimizations are performed of the function
\begin{align}\label{eq:LM_f}
	f(\LECvec, O, \lambda) = \chi^2(\LECvec) + \lambda O(\LECvec)
\end{align}
for various values of $\lambda$, known as the Lagrange multiplier.
Each such minimization results in a set of LECs, $\LECvec_\lambda$.
The obtained chi squared value, $\chi^2(\LECvec_\lambda) \equiv \chi_\lambda^2$,
is the minimum possible value,
under the constraint that $O = O(\LECvec_\lambda) \equiv O_\lambda$.
In this way all values of $O$ that are attainable under the constraint that
$\Delta\chi^2\leq L$ are found by varying $\lambda$, and so also the statistical uncertainty.
Note that the observable $O$ may or may not be part of the chi-squared function used to fit the LECs.

One potential complication with the minimization of $f$ is that
we do not know before the minimization what values of $\lambda$
that will produce a reasonable $\Delta\chi^2$, i.e.\ a change close to $L$.
To find reasonable $\lambda$ values, we can approximate $f$ using a Taylor expansion,
\begin{align}\begin{split}
	f(\LECvec_0 + \Delta\LECvec, O, \lambda) \approx& (\chi_0^2 + O_0) + \lambda j_0^T\Delta\LECvec\\
		&+ \frac{1}{2}(\Delta\LECvec)^T(H_0 + \lambda h_0)\Delta\LECvec.
\end{split}\end{align}
The LECs that minimize this approximate, quadratic expression, are given by
\begin{align}\label{eq:LM_approx_alpha}
    \Delta\LECvec_{\rm approx.} = -\lambda {(H_0+\lambda h_0)}^{-1}j_0
\end{align}
Inserting this into Eq.~\eqref{eq:chi2_taylor}, we get
\begin{align}\begin{split}\label{eq:LM_approx_chi2}
    \Delta\chi^2_{\rm approx.} &= \frac{1}{2}{(\Delta\LECvec_{\rm approx.})}^T H_0\Delta\LECvec_{\rm approx.}\\
    	&\approx \frac{1}{2}\lambda^2j_0^TH_0^{-1}j_0 = \frac{\lambda^2}{4L}\sigma_{\rm linear}^2,
\end{split}\end{align}
where the approximation assumes $H_0 + \lambda h_0 \approx H_0$ and
the last equality used Eq.~\eqref{eq:sigma_linear}.
Thus, to obtain an approximate deviation in $\chi^2$ equal to $\tilde{\lambda}^2L$, we need to use
\begin{align}\label{eq:LM_lambda_guess}
	\lambda = \tilde{\lambda}\frac{2L}{\sigma_{\rm linear}}.
\end{align}
This approximation may be inaccurate when the linear covariance approximation is insufficient.

It is possible to construct an \emph{approximate LM} method,
using Eqs.~\eqref{eq:chi2_taylor},~\eqref{eq:O_approx} and~\eqref{eq:LM_approx_alpha}.
This approximate LM method needs no minimizations of $f$, instead the first- and second-order derivatives
of $\chi^2$ and $O$ with respect to the LECs are needed.
On the other hand, the exact LM method requires no derivative information but instead minimizations of $f$.

\section{Results\label{sec:results}}

The six methods presented here, divided into covariance matrix methods and LM methods,
are described in Sec.~\ref{sec:method}.
To compare these methods, we have employed the so called NNLOsim potential with
$\Lambda = \unit[500]{MeV}$ and $T_{\rm lab}^{\max} = \unit[290]{MeV}$ from Ref.~\cite{carlsson2016}.

\begin{figure}
    \centering
    \includegraphics[width=\columnwidth]{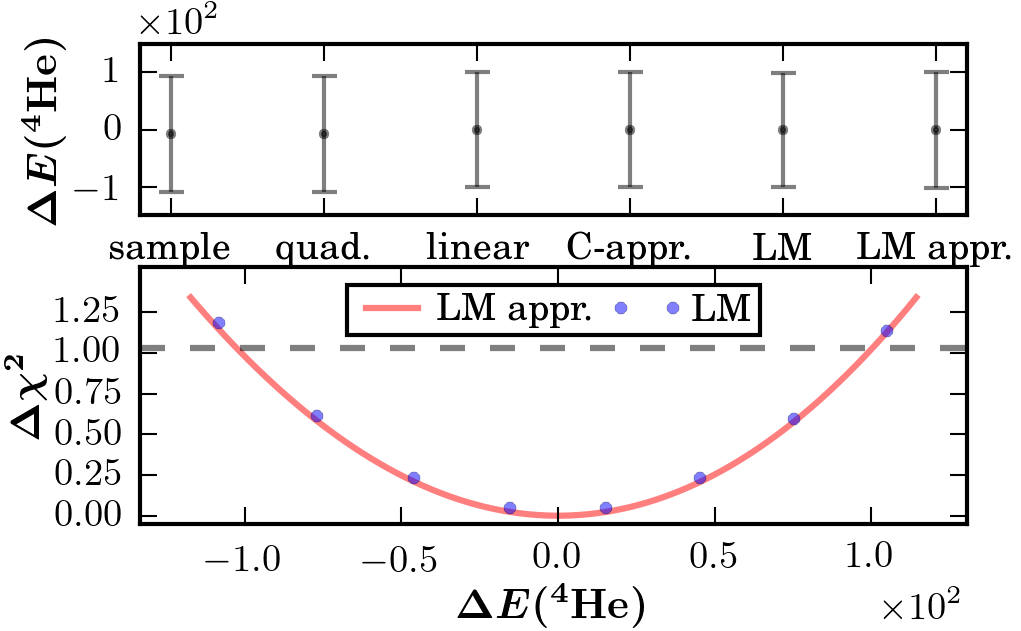}
    \caption{\label{fig:He4E}\co Upper: Propagated statistical uncertainties for
    	the helium-4 binding energy, in keV.
    	The methods, from left to right, are:
    	(i): Monte Carlo sampling using the covariance matrix,
    	(ii): as (i), using a quadratic approximation of the LEC dependence of the observable,
    	(iii): as (ii), using instead a linear approximation,
    	(iv): as (iii), except using an approximate covariance matrix,
    	(v): The LM method and
    	(vi): a quadratic approximation to (v).
    	See text for details.
    	Lower: The variation in $\chi^2$ as a function of the observable
    	value, as obtained using the LM method.
    	The solid line shows the quadratic approximation of the LM calculations.}
\end{figure}

\begin{figure}
    \centering
    \includegraphics[width=\columnwidth]{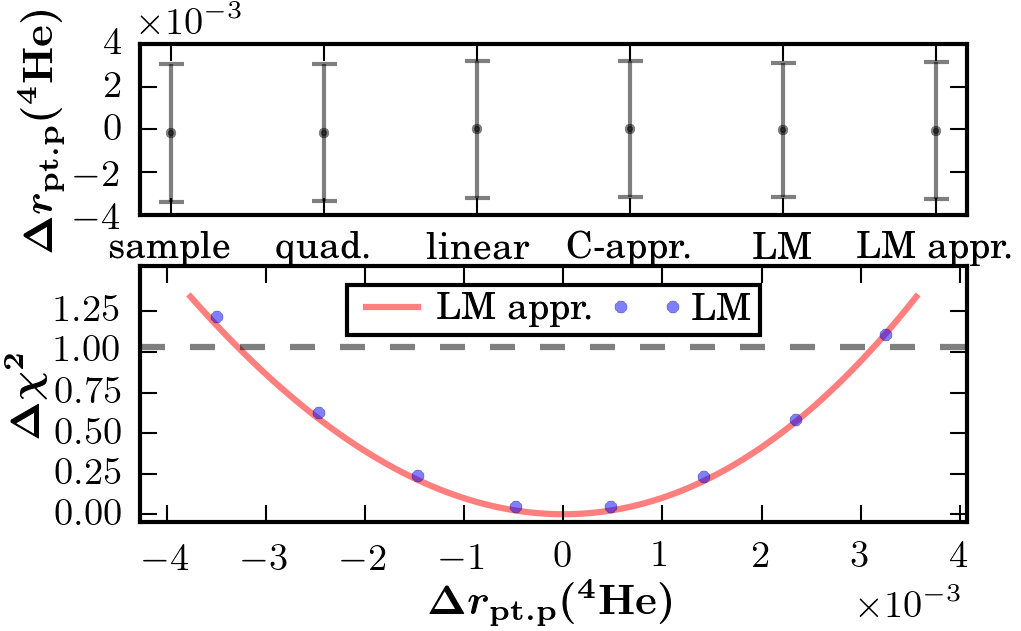}
    \caption{\label{fig:He4R}\co As Fig.~\ref{fig:He4E}, for the helium-4 point-proton radius, in fm.}
\end{figure}

\begin{figure}
    \centering
    \includegraphics[width=\columnwidth]{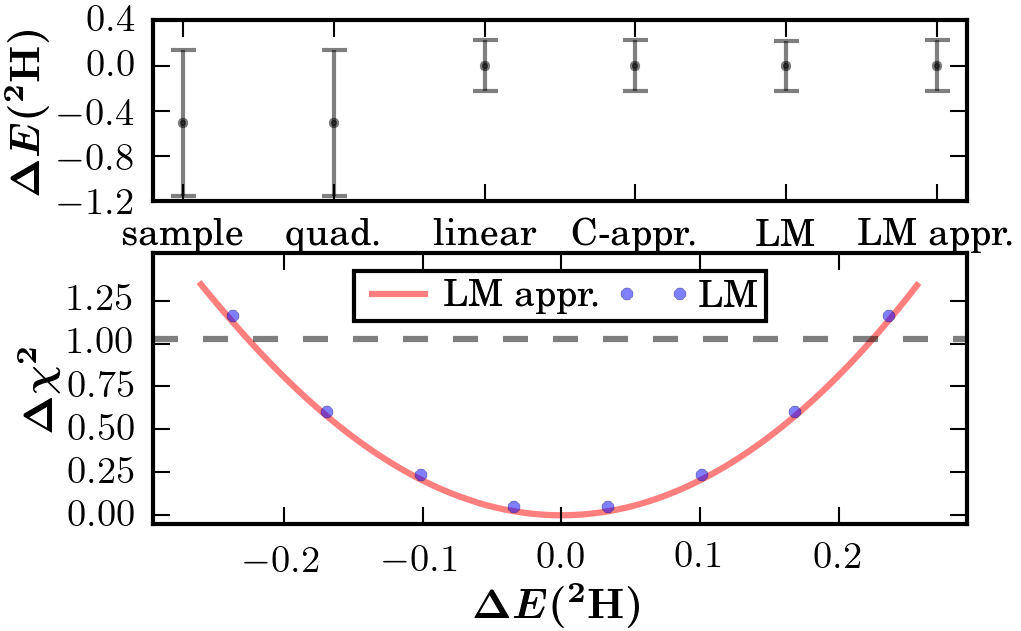}
    \caption{\label{fig:H2E}\co As Fig.~\ref{fig:He4E}, for the deuteron binding energy, in keV.}
\end{figure}

\begin{figure}
    \centering
    \includegraphics[width=\columnwidth]{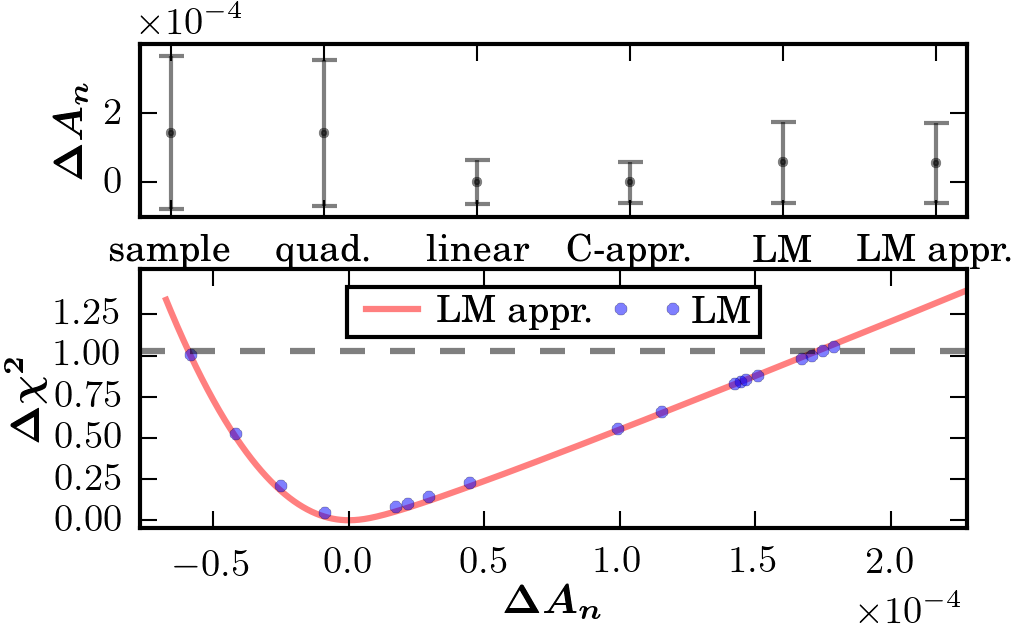}
    \caption{\label{fig:An}\co As Fig.~\ref{fig:He4E}, for the neutron-proton analyzing power at
    	laboratory scattering energy $\unit[175.26]{MeV}$ and center-of-mass scattering angle $97.63$
    	degrees.}
\end{figure}

We focus on four different observables:
The helium-4 binding energy $\OEa$ and point-proton radius $\Orpa$ in Figs.~\ref{fig:He4E} and~\ref{fig:He4R},
the deuteron binding energy $\OEd$ in Fig.~\ref{fig:H2E}
and finally the neutron-proton analyzing power $A_n$ at
laboratory scattering energy $\unit[175.26]{MeV}$, shown
for center-of-mass angle $\theta_{\rm c.m.} = 97.63$ degrees in Fig.~\ref{fig:An}.
$\OEd$ and $A_n$ are part of the chi-squared function that was used in the construction of NNLOsim,
while $\OEa$ and $\Orpa$ are predictions.
The upper panel of each figure contains a comparison between the obtained statistical uncertainties
of the methods.
Note that all derivatives have been calculated using automatic differentiation, except in the calculations
involving the approximate covariance matrix, $\tilde{C}$, where finite differences is used.
In the lower part of the figures, the function $\Delta\chi^2(\Delta O)$ is shown.
For $\OEa$ and $\Orpa$, all six methods result in the same statistical uncertainty while for
$\OEd$ and $A_n$ there are some discrepancies.


Just as for $\OEa$ and $\Orpa$, the statistical uncertainties produced by the various methods
agree for the vast majority of observables that we have looked at.
This includes $\pi$N and NN scattering data and ground-state properties of $A=2-4$ nuclei.
There are, however, some exceptions where non-linearities in the observables
with respect to the LECs cause discrepancies.

To explain the discrepancies in Fig.~\ref{fig:H2E} for $\OEd$,
we need to look at the eigen directions in the LEC space,
given by the covariance matrix.
It turns out that the linear uncertainty in $\OEd$ is almost entirely determined from one particular
eigen direction. In this case, this is also the direction picked up by the LM method.
Since there are almost no non-linearities in this particular direction, the LM method results
in the same uncertainty as the linear approximation.
In other directions, not picked up by the LM method,
$\OEd$ has an almost purely quadratic dependence on the LECs.
These quadratic variations are picked up by the MC sampling and the quadratic approximation, as they
consider variations in all directions in the LEC space.
This is a case where the MC sampling and the quadratic approximation produces
more accurate statistical uncertainties than the LM method.

For the analyzing power in Fig.~\ref{fig:An}, the situation is a bit different.
$A_n(\theta_{\rm c.m.})$ is sensitive mainly to one particular eigen direction in the LEC space.
However, as $\theta_{\rm c.m.}$ varies the derivative in that direction crosses zero, causing the
linear uncertainty to almost vanish.
This obtained statistical uncertainties for this particular angle is shown in Fig.~\ref{fig:An}.
Therefore, it is the mainly quadratic variations in the other directions
that contribute to the larger uncertainties obtained using the other methods.
In these cases, it is important to include the quadratic dependence of the observable on the LECs to
correctly capture the uncertainty.

\section{Discussion\label{sec:discussion}}

For the vast majority of observables, including $\OEa$ and $\Orpa$,
all methods agree in their determination of statistical uncertainties, as stated in Sec.~\ref{sec:results}.
This suggests that, within the uncertainty limits, the chi-squared surface is purely quadratic in the LECs
and the observables in question are linear in the LECs.
A deviation from a purely quadratic expression of the chi-squared function would not be detected by any
of the covariance based methods, as all of these rely on a quadratic approximation and ignores higher-order
terms.
The LM method, on the other hand, does not assume a quadratic form of the chi-squared function.
Therefore, an agreement between all methods would suggest that the quadratic approximation is valid.
Furthermore, an agreement between the results using the exact and the approximate covariance matrix would suggest
that it is safe to ignore the second-derivative term in Eq.~\eqref{eq:H_elem}.
The only difference between the quadratic and linear methods is the amount of terms used in the
Taylor expansion of the observable of interest. Thus, their identical statistical uncertainties indicate
that the quadratic term for the observable is negligible in these cases.

To check whether the chi-squared surface is quadratic around the minimum for the NNLOsim interaction,
we evaluated the chi-squared function in the eigen directions up to $\Delta\chi^2 = L$.
We found that in one direction there is a slight contribution
from higher-order terms.
We do not expect this to have a big influence on the analysis, which is also
suggested by the observed general agreement between the linear and LM methods.
An observed disagreement between the methods occurred primarily when an observable $O$ was
non-linear in the LECs around the minimum.

From these examples, a few conclusions can be drawn regarding the feasibility of these methods in this particular case,
\begin{itemize}
	\item The covariance matrix for the LECs is enough to capture
		  the statistical variations of the LECs.
	\item In some cases, a linear relationship between observable and LECs is not sufficient to correctly
		  capture the propagated statistical uncertainties.
	\item Only the MC sampling and the quadratic approximation are able to take into account
		  variations in all directions in the LEC space.
\end{itemize}

Note that, if the statistical uncertainties had been larger, more discrepancies between the methods
would be expected, as the various approximations used would no longer be valid.
To test this hypothesis, we simulated larger statistical uncertainties by scaling all uncertainties
$\sigma_n^{\rm (tot)}$ by a factor $\gamma = 10$. This is equivalent to changing the limit of the allowed
change in the chi-squared function, $L$, to $\gamma^2 L$.
In this extended range, the chi-squared surface is no longer quadratic, containing significant contributions
from higher-order terms.

\begin{figure}
    \centering
    \includegraphics[width=\columnwidth]{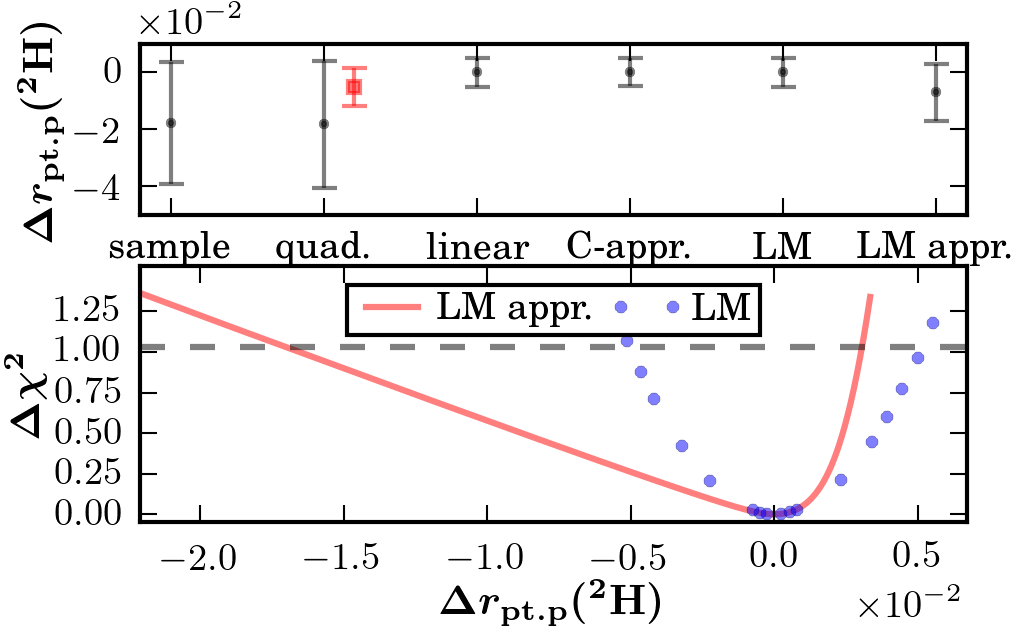}
    \caption{\label{fig:H2R_f10}\co As Fig.~\ref{fig:He4E}, for the deuteron point-proton radius in fm,
    	when using artificially enlarged uncertainties in the observables,
    	$\gamma\sigma_n^{\rm (tot)}$ with $\gamma = 10$.
    	This results in larger statistical uncertainties and a non-quadratic chi-squared surface around the
    	minimum. The error bar with a square
    	is a modified quadratic propagation to account for higher-order terms in the chi-squared surface,
    	see the text for details.}
\end{figure}

Using the original errors, $\sigma_n^{\rm (tot)}$, all methods produce equal uncertainties for $\Orpd$.
When instead using the errors $\gamma\sigma_n^{\rm (tot)}$, with $\gamma = 10$, the situation is
different, as shown in Fig.~\ref{fig:H2R_f10}.
The MC sampling and the quadratic approximation results in almost equal uncertainties.
This suggests that $\Orpd$ is still approximately quadratic in the LECs within this larger range.
The discrepancy between the exact and the approximate LM method is then due to
the higher-order terms in the chi-squared surface.
Since only the LM method is capable to account for a non-quadratic chi-squared surface, it has in this case
a distinct advantage over the other methods presented here.

The higher-order terms in the chi-squared surface can be approximately quantified by
calculating the uncertainties in the eigen directions through direct evaluations of the
chi-squared function in these directions.
One standard deviation, $\sigma_{\rm{calc},i}$, for the direct calculations
in eigen direction $i$ is defined by all $\Delta a$ such that
$\Delta\chi^2(\Delta a\bm{x}_i) \leq \gamma^2 L$ where $\bm{x}_i$ is eigen direction $i$.
The deviations due to higher-order terms are then defined by
the ratios $\sigma_{\rm{calc},i}/\sigma_{\rm{quad.},i}$ where $\sigma_{\rm{quad.},i}$ are the uncertainties
given by the quadratic approximation of the covariance matrix.

For the $26$ directions, nine directions have a deviation of more than $5$\% and the ratios for these
directions are in the range $0.3$ to $0.9$.
This indicates that the chi-squared surface tends to increase faster than what is estimated by the
second derivatives alone and the MC sample will tend to overestimate the uncertainty.
The error bar with a square in Fig.~\ref{fig:H2R_f10} is the same as the quadratic approximation
except that eigen values of the covariance matrix are taken from the explicit evaluations
of the chi-squared function mentioned above.
This is a crude way to account for higher-order terms but shows explicitly the influence of these terms.
A more sophisticated way to incorporate higher-order terms in the covariance matrix could be to calculate
the variances and covariances of the LECs using the LM method.

Apart from the estimated uncertainties, the covariance matrix method has the advantage over the LM method that
it does not require to minimize the objective function.
This is most important in cases where the objective function is expensive to calculate or not readily available.
Another issue with the LM method is that it is computationally challenging to calculate covariances between
observables, as two Lagrange multipliers must be used. Using the covariance matrix the propagated
covariances are straight forward to obtain.
However, when using the covariance based methods it is important to make sure the chi-squared surface is
quadratic around the minimum.

Therefore, the main conclusion of these investigations is that to propagate statistical uncertainties using
a chiral interaction the quadratic approximation using the covariance matrix is sufficient.
For this, only the interaction itself and an accompanying covariance matrix for the LECs
are needed. This is true as long as the LECs are well constrained by data, i.e.\ the statistical
uncertainties are small enough that the covariance matrix is enough to capture the variations in the LECs.
If this is not the case, the LM method is more accurate.

\begin{acknowledgments}
The author thanks C.~Forss\'en, A.~Ekstr\"om and M.~Hjorth-Jensen for
valuable discussions,
and acknowledges the hospitality of Oslo University, Norway.
The research leading to these results has received funding from
the European Research Council under the
European Community's Seventh Framework Programme (FP7/2007-2013) / ERC Grant No.\ 240603,
the Swedish Foundation for International Cooperation in Research and Higher Education
(STINT, Grant No.\ IG2012-5158),
the Swedish Research Council under Grant No.\ 2015-00225 and
the Marie Sklodowska Curie Actions, Cofund, Project INCA 600398.
Some computations were performed on resources at NSC in Link\"{o}ping
provided by the Swedish National Infrastructure for Computing.
\end{acknowledgments}

\bibliography{lamult}

\end{document}